\documentclass[10pt, twoside]{article}
\usepackage[cmex10]{amsmath}
\usepackage{amssymb}
\usepackage{mathrsfs}
\usepackage{amsthm}
\usepackage{soul}
\usepackage{graphicx}
\usepackage{hyperref}
\usepackage{cite}
\usepackage{fancyhdr}
\usepackage{makeidx}
\usepackage{authblk}
\usepackage{sectsty}
\sectionfont{\centering}
\subsectionfont{\centering}
\newtheorem{theo}{\bf Theorem}[section]
\newtheorem{pro}{Proposition}[section]
\newtheorem{cor}{Corollary}[section]
\newtheorem{lem}{Lemma}[section]
\newtheorem{exa}{Example}

\newcommand{\pr} {{\bf Proof. \hspace{0.5cm}}}
\newcommand{\m} {\mathfrak{m}}

\newcommand{\C} {\mathcal{C}}
\newcommand{\A} {\mathcal{A}}
\newcommand{\F} {\mathbb{F}}

\date{}
\begin{document}

\centerline{}

\centerline{}

\centerline {\Large{\bf Primitive Idempotents and Constacyclic Codes over Finite Chain Rings}}

\centerline{}

\newcommand{\mvec}[1]{\mbox{\bfseries\itshape #1}}

\centerline{\bf {Mohammed  Elhassani CHARKANI }}
\centerline{Department of Mathematics, Faculty of Sciences}
\centerline{Dhar-Mahraz-F$\grave{e}$s, Sidi Mohamed Ben Abdellah University}
\centerline{ Fez-Atlas, 30003, Morocco}
\vspace{0.5cm}
\centerline{\textbf{Jo\"el  KABORE \footnote{Corresponding author}}}
\centerline{Department of Mathematics}
\centerline{University Joseph Ki-Zerbo}
\centerline{Ouagadougou, Burkina-Faso}
\vspace{0.5cm}

{\bf Abstract.}
Let $R$ be a commutative local finite ring. In this paper, we
construct the complete set of pairwise orthogonal primitive
idempotents of $R[X]/<g>$ where $g$ is a regular polynomial in
$R[X]$. We use this set to decompose the ring $R[X]/<g>$ and to give
the structure of constacyclic codes over  finite chain rings. This
allows us to describe generators of the dual code $\mathcal{C}^\bot$
of a constacyclic code $\C$ and to characterize non-trivial
self-dual constacyclic codes over finite chain rings.
\vspace{0.5cm}


\section{Introduction}

Constacyclic codes over finite commutative rings are an important
class of linear block codes. Let $R$ be a commutative ring with
identity, it's well-known that for a given unit  $\lambda$, the
$\lambda$-constacyclic codes over $R$ are ideals of the ring
$R[X]/<X^n- \lambda>$.
 When studying constacyclic codes over finite chain rings, many authors assume that the code length is prime with  the characteristic of  its residue field.
 This ensures that the polynomial $X^n - \lambda$ have no multiple factor; in this case the codes are called simple root constacyclic codes, else they are called repeated root constacyclic codes. Simple root constacyclic codes have been extensively study by many authors \cite{ref35, ref36, ref37, ref44, ref46, ref70}.

 P. Kanwar and S. Lopez-Permouth gave the structure of cyclic codes over $\mathbb{Z}_{p^m}$, the ring of integers modulo $p^m$ \cite{ref35}. Q. Dinh and S. Lopez-Permouth extended this structure to
cyclic codes and negacyclic codes of odd length over finite chain ring \cite{ref46}.
They gave some necessary and sufficient conditions for the existence of non-trivial self-dual cyclic codes. E. Mart\'{i}nez-Moro and I. F. R\'{u}a generalized these results to multivariable codes over finite chain rings.
S.T.  Dougherty studied the  cyclic codes of arbitrary length over
 the ring of integers modulo $m$ \cite{ref55}.

    Using this results,   A. Batoul et al. considered the self-duality of cyclic codes over finite chain rings \cite{ref72}.
     Some additionally necessary and sufficient conditions for the existence of non-trivial negacyclic and cyclic self-dual codes are given in \cite{ref70}
     with a different method from that given in \cite{ref35, ref46}.\\

     The idempotents are   very excellent
tools to describe finitely generated modules over a decomposable
commutative ring $A\,=\,\prod_{i=1}^{n}A_{i}$. Indeed   if
$A\simeq\,\prod_{i=1}^{n}A_{i}$ is a decomposable ring then the
studying of the structure of  finitely generated modules over the ring
$A$ is reduced to studying the structure  of finitely generated
modules over each component  ring $A_{i}$. Idempotents have been
used intensively  to describe minimal cyclic codes  over finite
fields (see \cite{ref2} ).

In this paper, we use idempotents of the quotient ring $R[X]/<X^n-
\lambda>$ to determine the structure of constacyclic codes over
finite chain rings. Our method standardize the results of
\cite{ref35, ref46, ref70, ref72}. We first construct  a
complete set of primitive pairwise orthogonal idempotents of
$R[X]/<g>$, where $R$ is a commutative finite local ring and $g$ is
a regular polynomial in   $R[X]$.  We use this family of idempotents
to construct simple root constacyclic codes over finite chain rings.

We also investigate  the dual code $\mathcal{C}^\bot$ of a
constacyclic code $\C$ and characterize non-trivial
self-dual constacyclic codes over finite chain ring. We show that all non-trivial
constacyclic self-dual codes can be determined by non-trivial cyclic
or non-trivial negacyclic self-dual codes.

\section{Preliminaries}

Let $R$ be a finite local commutative ring, $\m$ be the maximal ideal of $R$ and $\F_q$ its residue field.
Let~ $\bar{}$~ be the natural surjective ring morphism given by:
$$
\begin{array}{l l l l}
\bar{}:& R &\longrightarrow & \F_q \\
       & r &\longmapsto & r + \m.
\end{array}
$$
This map extends naturally to a ring morphism from $R[X]$ to $\F_q[X]$ by sending $X$ to $X$.\\
An ideal $I$ in $R$ is primary if $I \neq R$ and whenever $xy \in
I,$ then either $x \in I$ or $y^n \in I$ for some positive integer
$n$. We say that two ideals $I$ and $J$ are coprime in $R$ if $I+ J= R$. A polynomial $f$ in $R[X]$ is called primary if $f R[X]$ is a primary ideal; regular if $f$ is not a zero divisor; basic irreducible if $\bar{f}$ is irreducible in $\F_q[X].$\\
Two polynomials $f, g \in R[X]$ are called coprime if $f R[X]$ and $g R[X]$ are coprime in $R[X]$; that is to say, there exists $u$ and $v$ in $R[X]$ such that $ f u+ g v= 1.$ This last relation is well-known as B\'ezout Identity.
Let Recall the Gauss Lemma which is an  additive property.

\begin{pro}[Gauss Lemma]\label{gauss}
Let $R$ be a commutative ring with identity. Let $f$ and $g$ be two
coprime polynomials in   $R[X]$. If $f$ divides the product $h g$ in
$R[X]$ then $f$ divides $h$ in $R[X]$.
\end{pro}
\pr Indeed, if $f$ and $g$ are two coprime polynomials in $R[X],$
then there exists $u$ and $v$ in $R[X]$ such that $fu+ gv=1.$ This
implies that $h= hfu+hgv$. Since $f$ divides $h g,$ then there
exists $w$ in $R[X]$ such that $hg=wf$ and $h= hfu+ wfv= f(hu +
wv)$. Therefore $f$ divides $h$ in   $R[X]$. \qed

\begin{pro}[\cite{ref38}, Theorem XIII.11]{~\newline}\label{pro1}
Let $R$ be a commutative local finite ring and $f$ be a regular polynomial in   $R[X]$. Then $f= \delta g_1...g_r$ where $\delta$ is a unit and $g_1, g_2, ..., g_r$ are regular
 primary pairwise-coprime polynomials.\\
Moreover, $g_1,..., g_r$ are unique in the sense that if $f=  \delta g_1...g_r = \beta h_1...h_s,$ where $\delta, \beta$ are units, and $\{g_i\},~ \{h_i\}$ are regular primary coprime polynomials, then $r =s,$ and after renumbering $g_i R[X] =h_i R[X],~ 1 \leq i \leq r.$
\end{pro}

The following result is very useful for determining coprime.
polynomials 
\begin{pro}[\cite{ref38}]\label{coprimcrit}
Let $R$ be a finite chain ring. Let $f$ and $g$ be two regular
polynomials in   $R[X]$. Then  $f$ and $g$ be two coprime
polynomials in $R[X]$ if and only if  $\overline{f}$ and
$\overline{g}$ be two coprime polynomials  in $k[X]$.
\end{pro}

The following result shows that we can reduce a study with regular polynomials to monic polynomials.

\begin{pro}[\cite{ref38}, Theorem XIII.6]\label{pro2}
Let $R$ be a commutative finite local ring and $f$ be a regular
polynomial in   $R[X]$. Then there is a monic polynomial $g$ with
$\bar{f}= \bar{g}$ and, for an element $a$ in $R$, $f(a)=0$ if and
only if $g(a)=0.$ Further, there is a unit $\delta$ in $R[X]$ with
$\delta f= g$.
\end{pro}

A code $\C$ of length $n$ over $R$ is nonempty subset of $R^n$; if in addition the code is a submodule of $R^n$, it is called linear code. In this paper all codes are assumed to be linear. For a given unit $\lambda \in R$, the $\lambda$-constacyclic shift $\sigma$ on $R^n$ is defined by $$\sigma(a_0,...,a_{n-1})=(\lambda a_{n-1},a_0,...,a_{n-2})$$ and a code of length $n$ over $R$ is said to be $\lambda$-constacyclic if it is invariant under the $\lambda$-constacyclic shift $\sigma$. Cyclic and negacyclic codes are examples of $\lambda$-constacyclic codes for $\lambda = 1$ and $-1$ respectively. The $\lambda$-constacyclic codes of length $n$ over $R$ are identified with ideals of $\frac{R[X]}{<X^n-\lambda>}$ by the identification:
$$ (a_0,a_1,...,a_{n-1})\longmapsto a_0+a_1x+...a_{n-1}x^{n-1};$$
where $x= X + <X^n-\lambda>$ is the equivalence class of $X$ in $\frac{R[X]}{<X^n-\lambda>}.$ \\
Given codewords $a=(a_0,a_1,...,a_{n-1})$, $b=(b_0,b_1,...,b_{n-1}) \in R^n$, their inner product is defined in the usual way: $$ a.b= a_0b_0+a_1b_1+...a_{n-1}b_{n-1},~\text{evaluated in}~ R.$$
The codewords $a$, $b$ are called orthogonal if $a.b=0$. The dual code $\mathcal{C}^\bot$ of $\C$ is the set of $n$-tuples over $R$ that are orthogonal to all codewords of $\C:$
$$\C^\bot=\{a~ |~ a.b=0,\forall a \in \C\}.$$
A code $\C$ is called self-orthogonal code if $\C \subseteq \C^{\bot}$ and self-dual code if $\C=\C^\bot$.\\

\begin{pro}[\cite{ref33}, Lemma 2.1]
Let $\lambda$ be a unit in $R$, the dual of a $\lambda$-constacyclic code is a $\lambda^{-1}$-constacyclic code.
\end{pro}

Let $f$ be the polynomial $f= a_0 +a_1 x+...+a_{n-1} x^{n-1} \in R[x]$, where $x=X+ <X^n- \lambda>$ and $a_0, a_1,...,a_{n-1} \in R$. The reciprocal polynomial of $f$  denoted by $f^*$ is defined as $f^*= a_0 x^{n-1}+a_{1}x^{n-2}+...+a_{n-1}$. Note that $(f^*)^*= f.$\\
The following result is easy to check.

\begin{pro}
Let $f$ and $g$ be two polynomials in $R[x]$ with $\deg f \geq \deg g.$ Then the followings hold:
\begin{itemize}
\item $(f + g)^* = f^* + x^{\deg f - \deg g} g^*;$
\item $(f g)^*= f^* g^*.$
\end{itemize}
\end{pro}

Let $\lambda$ be a unit in $R$ and $\C$ be an ideal of $R[X]/ <X^n- \lambda>.$ We define $\C^*$ by $\C^*=\{ f(x)^* \in R[x] : f(x) \in I\}.$
We let $$\A(\C)= \{g(x) \in R[x]: f(x)g(x)=0, ~\forall~ f(x) \in \C \}.$$
The set $\A(\C)$ is  an ideal of $R[X]/ <X^n- \lambda>$ called annihilator of $\C.$

\begin{pro}[\cite{ref33}, Proposition 2.3]\label{c11}{~\newline}
Let  $\lambda$ be a unit in $R$, $\C$ be a $\lambda$-constacyclic code of length $n$ over $R$ and $\C^{\bot}$ be the dual code of $\C.$ Then $$\C^{\bot}= \A(\C)^{*}.$$
\end{pro}

\section{The quotient ring \(R[X]/<g> \) and the idempotents}

Let $R$ be a commutative ring with identity. An element $e$ of $R$ is called idempotent if $e=e^2;$ two idempotents $e_1, e_2$ are said to be orthogonal if $e_1e_2=0.$ An idempotent of $R$ is said primitive if it is non-zero and cannot be written as sum of non-zero orthogonal idempotents.\\
A set $\{e_1,...,e_r\}$ of elements of $R$ is called a complete set of idempotents if $\sum_{i=1}^r e_i=1.$

If $\{e_1,...,e_r\}$ is a complete set of pairwise orthogonal idempotents of $R$, it's easy to show that $R = \oplus_{i=1}^r e_i R.$

\begin{pro}\label{primi}(\cite{ref88}, Proposition 22.1){~\newline}
Let $R$ be a commutative ring with identity. There exists at most
one complete set of pairwise orthogonal primitive idempotents
$\{e_1,...,e_r\}$ of $R$. Moreover, any idempotent in $R$ is
uniquely written as a finite sum of primitive idempotents of this
set.

\end{pro}
\pr
Let $\{e_1,...,e_r\}$ be a complete set of pairwise orthogonal primitive idempotents in $R$.
If $\theta$ is an idempotent in $R,$ then $1- \theta$ is also an idempotent in $R$ and we have: $1 = \theta + (1- \theta).$ This implies that $e_i= \theta e_i + (1- \theta)e_i.$ Since $e_i$ is primitive for all $i \in \{1,..,r\},$ then $\theta e_i=0$ or $\theta e_i = e_i.$ There exists $I \subseteq \{1,..,r\}$ such that $\theta=\sum_{i=1}^r \theta e_i= \sum_{i \in I } \theta e_i= \sum_{i \in I } e_i.$ Moreover, if $\theta$ is primitive, then there exists $i \in \{1,..,r\}$ such that $\theta = e_i,$ whence the set $\{e_1,...,e_r\}$ is unique.\\
We suppose that there exists $J \subseteq \{1,..,r\}$ such that $J \neq I$ and $\theta= \sum_{i \in I } e_i= \sum_{i \in J } e_i.$ Then, there exists $j \notin I \cap J$ such that $\theta e_j = e_j$ and  $\theta e_j=0,$ absurd.
\begin{flushright}
\qed
\end{flushright}

Let $R$ be a finite local commutative ring and $g$ be a regular
polynomial in   $R[X]$. From Propositions \ref{pro1} and \ref{pro2},
we can assume $g$ is a monic polynomial in $R[X]$ and  factors
uniquely as a product of monic primary pairwise coprime polynomials:
$g= \prod_{i=1}^r g_i.$ We let $\hat{g}_i=\frac{g}{g_i}.$ Note that
$g_i$ and $\hat{g}_i$ are coprime  and regular polynomials.

\begin{theo}\label{general}
Let  $R$ be a finite local commutative ring and $g$ be a monic polynomial in $R[X]$ such that $g = \prod_{i=1}^{r} g_i$ is the unique factorization of $g$ into
a product of monic primary pairwise coprime polynomials. Let $x = X + <g>$ be the equivalence class of $X$ in $R[X]/<g>.$
 The ring $R[X]/<g>$ admits a unique complete set of primitive pairwise orthogonal idempotents $\{e_1, e_2,...,e_r\}$ given by:
$$ e_i= v_i(x) \hat{g}_i(x),~\text{where}~ v_i(x) \in R[x].$$\\
Moreover $e_i R[x] \cong \frac{R[X]}{<g_i>}$ and $R[x]=\oplus_{i=1}^r e_i R[x].$
\end{theo}
\pr Let $g = \prod_{i=1}^{r} g_i$ be the unique factorization of $g$
into a product of monic primary pairwise coprime polynomials of $g$
in   $R[X]$. Since $g_i$ and $\hat{g}_i= g/g_i$  are  coprime in
$R[X]$, then there exists $u_i, v_i \in R[X]$ such that $u_i g_i +
v_i \hat{g}_i=1.$ We let $e_i= v_i(x) \hat{g}_i(x)$ where $x = X +
<g>$ is the equivalence class of $X$ in $R[X]/<g>.$ We have:
$$e_i^{2}= v_i(x) \hat{g}_i(x)(1- u_i(x)v_i(x))= v_i(x) \hat{g}_i(x)=e_i.$$
If $i \neq j,$ then $e_i e_j= v_i(x) \hat{g}_i(x)v_j(x) \hat{g}_j(x)=0.$  Hence $\{e_1, e_2,...,e_r\}$ is a set of   pairwise orthogonal idempotents.\\
The proposition \ref{gauss} (Gauss Lemma) ensures the uniqueness of
$e_i.$ Indeed, assume $(u_i^{'}, v_i^{'})$ is another pair of
polynomials in $R[X]$ such that: $u_i^{'}g_i+ v^{'}_i \hat{g}_i=1;$
then $ u_i^{'} g_i + v_i^{'} \hat{g}_i = u_i g_i + v_i \hat{g}_i,$
which gives $(u_i^{'}- u_i)g_i = (v_i-v_i^{'})\hat{g}_i.$
 Since $g_i$ and $\hat{g}_i$ are  coprime and regulars, then $g_i$ divides $v_i- v_i{'}$ from Gauss Lemma. Then there exists $h$ in $R[X]$ such that: $v_i-v_i^{'}= h g_i.$ Hence $v_i= h g_i+ v_i^{'},$ and $e_i = v_i(x)\hat{g}_i(x)= v_i^{'}(x)\hat{g}_i(x).$\\
Since $\hat{g}_1, \hat{g}_2,..., \hat{g}_r$ are  coprime, there exists $v_1, v_2,...,v_r \in R[X]$ such that $\sum_{i=1}^r v_i \hat{g}_i=1;$ hence $\sum_{i=1}^{r} e_i=1.$\\
Let
$$\begin{array}{l l l l}
T:& R[X] &\longrightarrow & e_i R[x]\\
& h & \longmapsto & e_i h= v_i(x)\hat{g_i}(x) h.
\end{array}
$$
$T$ is an onto ring homomorphism and by  the Gauss Lemma
(Proposition \ref{gauss}) we see that  $\ker T=< g_i>,$ and hence by
the first isomorphism theorem, we deduce $R[X]/<g_i> \cong e_i
R[x].$ Since $g_i$ is primary in  $R[X],$ then $R[X]/<g_i>$ is a
local ring, so it is an indecomposable ring. Therefore $\{e_1,
e_2,...,e_r\}$ is a set of primitive idempotents.

 \begin{flushright}
\qed
\end{flushright}

\section{Constacyclic codes over finite chain ring}

A finite chain ring is a finite commutative ring with identity such that its ideals are linearly ordered by inclusion. The following result is well know and characterizes finite chain rings.

\begin{pro}[\cite{ref46}, Proposition 2.1]
Let $R$ be a finite commutative ring with identity, the following conditions are equivalent:
\begin{enumerate}
\item $R$ is a local ring and the maximal ideal of $R$ is principal,
\item $R$ is a local principal ideal ring,
\item $R$ is a chain ring.
\end{enumerate}
\end{pro}

If $R$ is a finite chain ring with maximal ideal $\gamma R;$ then $\gamma$ is nilpotent with nilpotency index some integer $t$ and the ideals of $R$ form the following chain:
$$ 0 = \gamma^{t} R \subsetneq \gamma^{t-1} R \subsetneq... \subsetneq {\gamma} R \subsetneq R.$$
We denote the residue field $R/ <\gamma>$ by $\F_{p^r}.$\\
It's well-known that for linear codes of length $n$ over a finite chain ring $R$, $|\C||\C^\bot|=|R|^n$ (see \cite{ref37}).

\begin{lem}[\cite{ref46}, Lemma 3.1] \label{prince} ~\newline
Let $R$ be a finite chain ring with maximal ideal $\gamma R,$ index of nilpotency $t$ and residue field $\F_{q}.$
Let $f$ be a monic basic irreducible polynomial in the ring $R[X]$ and $x= X+ <f>$ be the equivalence class of $X$ in $\frac{R[X]}{<f>}.$ Then $\frac{R[X]}{<f>}$ is a finite chain ring with maximal ideal $\gamma R[x]$ and index of nilpotency $t$.
\end{lem}

Since $(n, p)= 1,$ the polynomial $X^n - \lambda$ factors uniquely as a product of monic basic irreducible pairwise coprime polynomials in $R$ (\cite{ref46}, Proposition 2.7). In the rest of paper we denote by $x= X+ <X^n - \lambda>$ the equivalence class of $X$ in $R[X] /<X^n - \lambda>,$ thus $R[X] /<X^n - \lambda>= R[x].$

\begin{theo}
Let $R$ be a finite chain ring with maximal ideal $\gamma R,$ index of nilpotency $t$ and residue field $\F_{q}.$ Let $\lambda$ be a unit in $R$, $X^n-\lambda= f_1 f_2...f_r$ be the unique decomposition of $X^n-\lambda$ into  product of monic basic irreducible pairwise coprime polynomials and $\{e_1,...,e_r\}$ be the complete set of primitive pairwise orthogonal idempotents in $R[X] /<X^n - \lambda>=R[x].$\\
Let $\C$ be a $\lambda$-constacyclic code of length $n$ over $R$. Then there exists a unique sequence of integers $(s_1,...,s_r)$ such that $0 \leq s_i \leq t$ and
$$\C= \oplus_{i=1}^r \gamma^{s_i} e_i R[x].$$
\end{theo}
\pr
 Since $R[x]= \oplus_{i=1}^r e_i R[x];$ then any ideal $I$ in $R[x]$ is written in the form $I= \oplus_{i=1}^r I_i,$ where $I_i$ is an ideal of $e_i R[x].$ By Theorem \ref{general}, we have $ e_i R[x] \cong R[X]/<f_i>.$ From previous lemma, we know that ideals of $R[X]/<f_i>$ are in the form $\gamma^j (R[X]/<f_i>),~ 0 \leq j \leq t;$ therefore $I_i =\gamma^j e_i R[x],~0 \leq j \leq t.$
\begin{flushright}
\qed
\end{flushright}

\begin{theo}\label{joe1}
Let $R$ be a finite chain ring with maximal ideal $\gamma R,$ index of nilpotency $t$ and residue field $\F_{q}.$  Let $\lambda$ be a unit in $R$ and $\C$ be a $\lambda$-constacyclic code of length $n$ over $R$. Then there exists a complete set of pairwise orthogonal idempotents $\{\theta_0,...,\theta_l\}$ in $R[X] /<X^n - \lambda>= R[x]$ such that:

$$\C= \oplus_{i=0}^{l-1} \gamma^{r_i} \theta_i R[x];$$
with $0 \leq r_0 < r_1 <...<r_{l-1}< r_l= t$ and $\sum_{i=0}^l \theta_i=1.$

Moreover there exists a unique family of pairwise coprime polynomials $g_0, g_1,...,g_l$ in $R[X]$  such that:\\

$\theta_i R[x] \cong R[X]/<g_i>, ~ \forall~ i \in \{0,1,...,l\}$ et $ \prod_{i=0}^l g_i= X^n- \lambda.$
 \end{theo}
\pr
Let $X^n-\lambda= f_1 f_2...f_r$ be the decomposition of $X^n-\lambda$ into product of monic basic irreducible pairwise coprime polynomials in $R$ and $\{e_1,...,e_r\}$ be the complete set of primitive pairwise orthogonal idempotents of $R[X] /<X^n - \lambda>=R[x].$\\ From the previous theorem:
$\C= \oplus_{i=1}^r \gamma^{s_i} e_i R[x],~ 0 \leq s_i \leq t.$ By reordering if necessary according to the powers of $\gamma$, we can write  $\C$ in the form:
\[
 \C= \bigoplus_{j~|s_j = r_0} \gamma^{r_0} e_j R[x] \bigoplus_{j~|s_j = r_1} \gamma^{r_1} e_j R[x] \bigoplus...\bigoplus_{j~|s_j = r_{l-1}} \gamma^{r_{l-1}} e_j R[x]
\]

with $0 \leq r_1 < r_2 <...<r_l= t.$ We let $\theta_i=
\sum_{j~|s_j=r_i} e_j,~ \forall i \in \{0,..., l-1\}$ and
$\theta_l=1\,-\,\sum_{i=0} ^{l-1} \theta_i$. Therefore, the set
$\{\theta_0, \theta_1,..., \theta_l\}$ is a complete set of pairwise
orthogonal idempotents; by construction this set is unique. We have:
$$\C= \oplus_{i=0}^{l-1} \gamma^{r_i} \theta_i R[x].$$

Since $e_j R[x] \cong \frac{R[X]}{<f_j>},~\forall~ 1 \leq j \leq r,$ then $\theta_i R[x]\cong \prod_{j~|s_j=r_i} \frac{R[X]}{<f_j>} \cong  \frac{R[X]}{<\prod_{j~|s_j=r_i} f_j>},$ by the Chinese Remainder Theorem.
We let $g_i= \prod_{j~|s_j=r_i} f_j,~ \forall~ 0 \leq i \leq l.$ It is clear that  $\prod_{i=0}^l g_i =X^n -\lambda.$
\begin{flushright}
\qed
\end{flushright}

\begin{cor}
Under the same assumptions as the Theorem \ref{joe1}, let $\C$ be a $\lambda$-constacyclic code of length $n$ over $R$. Then

$$\C= (\oplus_{i=0}^{l-1}\gamma^{r_i} \theta_i) R[x].$$
\end{cor}
\pr From previous theorem, we have: $\C= \oplus_{i=0}^{l-1}
\gamma^{r_i} \theta_i R[x]$ with $0 \leq r_0 < r_1 <...<r_l= t.$ We
let $w= \sum_{i=0}^{l-1} \gamma^{r_i} \theta_i.$ It's clear that $w
R[x] \subseteq \C.$ Reciprocally, if $b \in \C$, then
$b=\sum_{i=0}^{l-1} \gamma^{r_i} \theta_i b_i$ with $b_i \in R[x],
~\forall~ 0 \leq i \leq l-1.$ For any idempotent $\theta_j \in R[x]$,
we have: $\theta_j b= \gamma^{r_j} \theta_j b_j= \theta_j w b_j.$
Therefore $b =\sum_{j=0}^{l-1} \theta_j b= \sum_{j=0}^{l-1} \theta_j
w b_j =( \sum_{j=0}^{l-1} \theta_j b_j) w;$ hence  $b \in w R[x].$
\begin{flushright}
\qed
\end{flushright}

\begin{cor}\label{card1}
Under the same assumptions as the Theorem \ref{joe1}, let $\C$ be a  $\lambda$-constacyclic code of length $n$ over $R$ such that
$$\C= \oplus_{i=0}^{l-1} \gamma^{r_i} \theta_i R[x]$$ with
$0 \leq r_0 < r_1 <...<r_l= t.$ Then:

$$|\C|= |\F_q|^{\sum_{i=0}^{l-1}(t- r_i) \deg g_i}.$$

\end{cor}
\pr
Since $\theta_i R[x] \cong R[X]/<g_i>$ then $$|\gamma^{r_i} \theta_i R[x]|=|\gamma^{r_i}(R[X]/<g_i>)|.$$
We let $A_i=R[X]/<g_i>.$
The map
$$
\begin{array}{l l l l}
\phi_i:& A_i& \longrightarrow & \gamma^{r_i}A_i\\
& h &\longmapsto& \gamma^{r_i} h
\end{array}
$$ is an epimorphism  and $\ker \phi_i = \gamma^{t-r_i}A_i.$ By the first isomorphism theorem $A_i/(\gamma^{t-r_i} A_i) \cong \gamma^{r_i} A_i.$ But  $A_i/(\gamma^{t-r_i} A_i) \cong R_i[X]/<\widetilde{g}_i>,$ where $R_i = R/<\gamma^{t-r_i}>$ and $\widetilde{g}_i = g_i + <\gamma^{t-r_i}>.$ Therefore:
$$
\begin{array}{l l}
|\gamma^{r_i} A_i|& =|A_i/(\gamma^{t-r_i} A_i)|= |R_i[X]/<\widetilde{g}_i>|=|R_i|^{\deg g_i}\\
& = (\frac{|R|}{|\gamma^{t-r_i} R|})^{\deg g_i}= |\F_q|^{(t-r_i) \deg g_i}.
\end{array}
$$

We deduce:  $$|\C|= \prod_{i=0}^{l-1} |\gamma^{r_i} \theta_i R[x]|= |\F_q|^{\sum_{i=0}^{l-1} (t- r_i) \deg g_i}.$$
\begin{flushright}
\qed
\end{flushright}

\begin{lem}
Let $R$ be a commutative ring.
\begin{itemize}
\item[$i)$] If $e_1$ et $e_2$ are orthogonal idempotents in $R[X]$ then $(e_1+ e_2)^*= e_1^* +e_2^*.$
\item[$ii)$] If $e$ is a primitive idempotent in $R[X]$, then $e_1^*$ is a primitive idempotent in $R[X]$.
\end {itemize}
\end{lem}
\pr
\begin{itemize}
\item[$i)$] If $e_1$ et $e_2$ are orthogonal idempotents in $R[X]$, then $e= e_1 +e_2$ is also an idempotent. Since $e e_i= e_i,$ for all $i \in \{1, 2\}$ we have $(e e_i)^*=e^* e_i^*= e_i^*,$ for all $i \in \{1, 2\}.$ Then $e^*$ is written in the form: $e^*= e_1^* + e_2^* + \theta$ where $e_1^*, e_2^*, \theta$ are pairwise orthogonal idempotents. Likewise  $$ e_1+ e_2= e =(e^*)^*= (e_1^*)^* + (e_2^*)^* + \theta^* + \beta = e_1 + e_2 + \theta^* + \beta,$$ where $e_1, e_2, \theta^*, \beta$ are pairwise orthogonal idempotents. We deduce $\theta^* + \beta= \theta^*= \beta=0;$ whence $(e_1+ e_2)^*= e_1^* +e_2^*.$
\item[$ii)$] It's obvious from $i).$
\end{itemize}
\begin{flushright}
\qed
\end{flushright}

\begin{lem}
Let $I$ be an ideal of $R[x]$ such that $I = \oplus_{1 \leq i \leq r}~ h_i R[x],$ then $I^{*}= \oplus_{1 \leq i \leq r}~ h_i^* R[x].$
\end{lem}
\pr Let $I$ be an ideal of $R[X]/<X^n -\lambda>=R[x]$ such that $I=
h_1 R[x] + h_2 R[x];$ it is clear that $I^*= h_1^* R[x] + h_2^*
R[x].$ Let $f \in h_1^* R[x]~ \cap~ h_2^* R[x],$ then $f = h_1^* u=
h_2^* v$ with $u, v  \in R[x].$ If  $f$ is non zero then  $f^*= h_1
u^* = h_2 v^*$.  This implies that  $f^* \in h_1 R[x] \cap h_2 R[x]$
and hence we deduce that $f^*=0$. We deduce that $h_1^* R[x] \cap
h_2^* R[x]= \{0\}.$
\begin{flushright}
\qed
\end{flushright}

\begin{theo}\label{joe2}
Under the same assumptions as the Theorem \ref{joe1}, let $\C$ be a  $\lambda$-constacyclic code of length $n$ over $R$ such that
$$\C= \oplus_{i=0}^{l-1} \gamma^{r_i} \theta_i R[x],$$ with
$0 \leq r_0 < r_1 <...<r_l= t.$ Then:

$$\C^{\bot}= \oplus_{i=0}^l \gamma^{t-r_i} \theta_i^{*} R[x].$$
\end{theo}
\pr
Let $D= \oplus_{i=0}^l \gamma^{t-r_i} \theta_i R[x].$
For all $i, j \in \{0,...,l\},$ we have:\\
$(\gamma^{r_i} \theta_i)(\gamma^{t-r_j}\theta_j)=0,$ then $D \subseteq \A(\C).$

From Corollary \ref{card1}, $|D|=|\F_q|^{\sum_{i=0}^l r_i \deg g_i}.$
We recall that $|\C||\C^\bot|=|R|^n$ ( see \cite{ref37}).
Then:
$$
\begin{array}{l l}
|\C^{\bot}|&= \frac{|R|^n}{|\C|}=|\F_q|^{nt-\sum_{i=0}^{l-1}(t-r_i) \deg g_i}\\
& = |\F_q|^{nt- \sum_{i=0}^{l-1} t \deg g_i+ \sum_{i=0}^{l-1} r_i\deg g_i}\\
&= |\F_q|^{t \deg g_l + \sum_{i=0}^{l-1} r_i\deg g_i}\\
\end{array}
$$
 Therefore:
$|\A(\C)|= |\A(\C)^*|=|\C^{\bot}|=|D|;$ whence $D =\A(\C).$
We conclude that $$\C^{\bot}= D^*= \sum_{i=0}^l \gamma^{t-r_i} \theta_i^{*} R[x].$$

Let $\lceil \frac{t}{2} \rceil$ be the smallest integer greater than or equal to $t/2.$ If $\C$ is a linear code over $R$ such that $\C \subseteq \gamma^{\lceil \frac{t}{2} \rceil} R^n,$ it is easy to see that $\C \subseteq \C^{\bot}.$ These codes are called trivial self-orthogonal codes. Moreover, if $t$ is even, then the code $\C = \gamma^{t/2} R^n$ is self-dual and called trivial self-dual code.

Let $\C \subseteq R^n$ be a linear code. The submodule quotient of $\C$ by $r \in R$ is a linear code defined by $$(\C:r)=\{a \in R^n : r a \in \C\}.$$
We have the following tower of linear codes over $R$ $$\C=(\C: \gamma^0) \subseteq ...\subseteq (\C: \gamma^{t-1})$$ and its projection to $\F_{p^r}$
$$\overline{\C}=\overline{(\C: \gamma^0)} \subseteq ...\subseteq \overline{(\C: \gamma^{t-1})}.$$

For a unit $\lambda \in R,$ note that if $\C$ is a $\lambda$-constacyclic code over $R$, then $(\C:\gamma^{i})$ is a $\lambda$-constacyclic code over $R$ and $\overline{(\C:\gamma^{i})}$ is a $\overline{\lambda}$-constacyclic code over $\F_{p^r},$ for $i \in \{0,1,..., t-1\}.$

The following result generalises Lemma 3.3 in \cite{ref33} to finite
chain rings.

\begin{pro}\label{p1.1}
Let $R$ be a finite chain ring with maximal ideal $<\gamma>,$ index of nilpotency $t$ and residue field $\F_{q}.$ Let $\lambda$ be a unit in $R$ and $\C$ be a non-trivial $\lambda$-constacyclic self-orthogonal code over $R$. Then $\overline{\lambda}= \pm 1.$
\end{pro}
\pr
We suppose $\C$ is a nontrivial $\lambda$-constacyclic self-orthogonal code over $R$.
If  $\overline{\C} \neq \{0\},$ then $\overline{\C}$ is a $\overline{\lambda}$-constacyclic self-orthogonal code  over $\F_q.$ It is well-known that the only constacyclic self-orthogonal codes over a finite field are cyclic and negacyclic codes(\cite{ref33}, Proposition 2.4);
whence $\overline{\lambda}= \pm 1.$

If $\overline{\C} = \{0\},$ then there exists a smallest positive integer $i$ with $1 \leq i \leq e-1$ such that any codeword $c \in \C$ can be written as: $c=\gamma^{i}a,$ with $a \in R^n.$ Without loss of generality, we can suppose $\C \subseteq <\gamma^{i}>.$ Since $\C$ is a non-trivial $\lambda$-constacyclic self-orthogonal code over $R$, then $i < \lceil \frac{e}{2} \rceil,$ that is to say $2i < e$ and $\overline{(\C:\gamma^{i})}$  is self-orthogonal. Indeed if $a, b \in (\C:\gamma^{i}),$ then $c_1=\gamma^{i} a$ and $c_2=\gamma^{i} b$ verify $c_1.c_2=\gamma^{2i}(a.b)=0;$ hence $a.b=0.$ Then $\overline{(\C:\gamma^{i})}$ is self-orthogonal over $\F_q$ and
$\bar{\lambda}=\pm 1.$
\begin{flushright}
\qed
\end{flushright}

The following result shows us there exists a one-to-one correspondence between cyclic codes (respectively negacyclic codes) and $(1+ \gamma^i \beta)$-constacyclic codes (respectively $(1+ \gamma^i \beta)$-constacyclic codes) over $R$, with $\beta \in R.$

\begin{pro}[\cite{ref78}, Corollary 4.5]{~\newline}\label{ken}
Let $R$ be a finite chain ring with maximal ideal $\gamma R,$ index of nilpotency $t$ and residue field $\F_{q}.$ Let $n$ be a positive integer such that $(n, q)=1,$  $\lambda \in 1+ \gamma R$ and $\beta  \in -1+ \gamma R.$ Then there exists a ring isomorphism between $R[X]/<X^n -1>$ (respectively  $R[X]/<X^n +1>$) and $R[X]/<X^n - \lambda>$ ( respectively $R[X]/<X^n - \beta>$).
\end{pro}

From Proposition \ref{p1.1} and Proposition \ref{ken}, we can reduce the study of non-trivial constacyclic self-dual codes over $R$ to non-trivial cyclic and negacyclic self-dual codes over $R$.

\section{Self-dual cyclic codes}

\begin{theo}\label{joe3}
Under the same assumptions as the Theorem \ref{joe1}, let $\C$ be a  $\lambda$-constacyclic code of length $n$ over $R$ such that
$$\C= \oplus_{i=0}^{l-1} \gamma^{r_i} \theta_i R[x]$$ with
$0 \leq r_0 < r_1 <...<r_l= t.$ Then $\C$ is a non-trivial self-dual code if and only if $\theta_i$ and  $\theta_j^{*}$ are associated  and $r_i+ r_j= t$, for all $i, j \in \{0,...,l-1\}$ such that $i+ j \equiv 0 \mod l-1$.
\end{theo}
\pr
If $\C= \oplus_{i=0}^{l-1} \gamma^{r_i} \theta_i R[x],$ then by Theorem \ref{joe2},  $$\C^{\bot}=\sum_{i=0}^l \gamma^{t-r_i} \theta_i^{*} R[x].$$ If $\C$ is self-dual we must have $\theta_l= 0.$ In this case
$\C^{\bot}=\sum_{i=0}^{l-1} \gamma^{t-r_i} \theta_i^{*} R[x]$ with $\sum_{i=0}^{l-1} \theta_i =1$ and $0 \leq r_0 < r_1 <...< r_{l-1}<t.$
We obtain the result by comparing $\gamma$ exponents.
\begin{flushright}
\qed
\end{flushright}

\begin{cor}
Under the same assumptions as the Theorem \ref{joe1}, let $\C$ be a cyclic code of length $n$ over $R$ such that
$$\C= \oplus_{i=0}^{l-1} \gamma^{r_i} \theta_i R[x]$$ with
$0 \leq r_0 < r_1 <...<r_l= t.$
If there exists a non-trivial cyclic self-dual code over $R$, then $t$ is necessary even.
\end{cor}
\pr If $\C$ is self-dual, then by Theorem \ref{joe3},  $\C=
\oplus_{i=0}^{l-1} \gamma^{r_i} \theta_i R[x],$ with $0 \leq r_0 <
r_1 <...<r_{l-1}< t.$ Let  $X^n-1= \prod_{i \in I} f_i$ be the
decomposition of $X^n-1$ into a product of monic basic irreducible
pairwise coprime polynomials in   $R[X]$. Let $\{e_i\}_{i \in I}$ be
the complete set of primitive pairwise orthogonal idempotents of
$R[X]/<X^n-1>= R[x]$ given in Theorem \ref{general}; For $i \in I$, there exists
$u_i \in R[x]$ such that $e_i = u_i(x)\hat{f_i}(x)$.
Let $\theta_{i_0} \in \{\theta_0,..., \theta_{l-1}\}$ the idempotent containing $e_0,$ that is to say $\theta_{i_0}= e_0 + \beta$ where $\beta$ is an  idempotent orthogonal to $e_0.$\\
Since that $f_0= X-1,$ and $e_0$ is unique, we have $$e_0^{*}= u_0^*(x) \hat{f_0}^*(x)=-x^{n-2}u_0^*(x)\hat{f_0}= \eta e_0$$ where $\eta$ is inversible in  $R[x]$.  Hence $\theta_{i_0}^*= e_0^{*} + \beta^* = \eta (e_0 + \mu \beta^*) = \eta \theta_{i_0}$ where $\eta \mu= 1$ in $R[x]$.

Let $i_1 \in \{0,...,l-1\}$ such that $i_1 + i_0 \equiv 0 \mod l-1.$ If $\C$ is self-dual then $\theta_{i_1}$ and $\theta_{i_0}^*$ are associated, hence $\theta_{i_1}$ and $\theta_{i_0}$ are associated. This gives $i_1 = i_0$ and $2r_{i_0}=t,$ whence $t$ is even.
\begin{flushright}
\qed
\end{flushright}

\begin{theo} \label{similar}
Under the same assumptions as the Theorem \ref{joe1}, let $\C$ be a cyclic code of length $n$ over $R$ with even index of nilpotency $t$ such that
$$\C= \oplus_{i=0}^{l-1} \gamma^{r_i} \theta_i R[x]$$ with
$0 \leq r_0 < r_1 <...<r_l= t.$

Then there exists a non-trivial cyclic self-dual code over $R$ if and only if there exists an idempotent $\theta_i \in \{\theta_0,..., \theta_{l-1}\}$ such that $\theta_i$ and $\theta_i^*$ are not associated.
\end{theo}
\pr
Assume that there exists $\theta_i \in \{\theta_0,..., \theta_{l-1}\}$ such that $\theta_i$ and $\theta_i^*$ are not associated. We have $1 + x^{n-1}= \sum_{j=0}^{l-1} \theta_j + \sum_{j=0}^{l-1} \theta_j^{*}= \theta_i + \theta_i^{*}+ \beta$, with $\beta = 1+ x^{n-1} - \theta_i - \theta_i^{*}$. Note that $\beta^{*} = \beta$. Let
$$\C= \gamma^{t/2-1} \theta_i R[x] \oplus \gamma^{t/2} \beta R[x] \oplus \gamma^{t/2+1} \theta_i^{*} R[x].$$
From Theorem \ref{joe2}, we deduce that $\C$ is self-dual.\\
Reciprocally, let $\C$ be a non-trivial self-dual cyclic code such that $\C= \oplus_{i=0}^{l-1} \gamma^{r_i} \theta_i R[x],$ with $0 \leq r_0 < r_1 <...r_{l-1}< t.$ Assume that for all $i \in \{0,...,l-1\}$, $\theta_i$ and $\theta_i^{*}$ are associated. Then by Theorem \ref{joe3}, we must have $r_i= t/2, ~\forall~ 0 \leq i \leq l-1.$ Then $\C$ is thus written in the form:
$\C=\gamma^{t/2} \oplus_{i=0}^{l-1} \theta_i R[x],$ which is absurd, since $\C$ is assumed to be non-trivial self-dual code.
\begin{flushright}
\qed
\end{flushright}

\begin{exa}
We give a non-trivial cyclic self-dual code of length $6$ over $\mathbb{Z}_{7^2}$. 

 Let $x= X +<X^6-1>$. The irreducible factors of $X^6 - 1$ over $\mathbb{Z}_7$ are:
$f_0 = X-1; f_1=X-3; f_2=X-2; f_3=X-6; f_4= X-4; f_5= X-5$ and the complete set of primitive pairwise orthogonal idempotents of $\mathbb{Z}_{7}[X]/<X^6-1>$ is given by:
$$
\begin{array}{l l}
\theta_{0}=& 6 (x^5+ x^4+x^3+x^2+x+1);\\
\theta_{1}=& 4x^5+5x^4+x^3+3x^2+2x+6;\\
\theta_{2}=& 5x^5+3x^4+6x^3+5x^2+3x+6;\\
\theta_{3}=&x^5+6x^4+x^3+6x^2+x+6;\\
\theta_{4}=&3x^5+5x^4+6x^3+3x^2+5x+6;\\
\theta_{5}=&2x^5+3x^4+x^3+5x^2+4x+6.\\
\end{array}
$$
From Theorem $5.4$ of \cite{ref35}, we deduce the complete set of primitive pairwise orthogonal idempotents of $\mathbb{Z}_{7^2}[X]/<X^6-1>$:
$$
\begin{array}{l l l}
e_0=& \theta_0^7=& 41(x^5+ x^4+x^3+x^2+x+1);\\
e_1=& \theta_1^7=& 46x^5+ 5x^4+8x^3+3x^2+44x+41;\\
e_2=& \theta_2^7=& 5x^5+ 3x^4+41x^3+5x^2+3x+41;\\
e_3=& \theta_3^7=& 8x^5+41 x^4+8x^3+41x^2+8x+41;\\
e_4=& \theta_4^7=& 3x^5+ 5x^4+41x^3+3x^2+5x+41;\\
e_5=& \theta_5^7=& 44x^5+ 3x^4+8x^3+5x^2+46x+41.\\
\end{array}
$$
This gives:
$$
\begin{array}{l l l}
e_0^{*}=& 41(x^5+ x^4+x^3+x^2+x+1)=&e_0;\\
e_1^{*}=& 41x^5+ 44x^4+3x^3+8x^2+5x+46=& 31e_5;\\
e_2^{*}=& 41x^5+ 3x^4+5x^3+41x^2+3x+5=& 30e_4;\\
e_3^{*}=& 41x^5+8 x^4+41x^3+8x^2+41x+8=& 48e_3;\\
e_4^{*}=& 41x^5+ 5x^4+3x^3+41x^2+5x+3=& 18 e_2;\\
e_5^{*}=& 41x^5+ 46x^4+5x^3+8x^2+3x+44=& 19 e_1.\\
\end{array}
$$
We let $\beta= 1+x^5-e_2-e_2^*= 4x^5+43x^4+3x^3+3x^2+43x+4$.
It's clear that $\beta^{*}=\beta$. By the previous theorem, we have the following self-dual cyclic code
$$\mathcal{C}= e_2 \mathbb{Z}_{49}[x] \oplus 7\beta \mathbb{Z}_{49}[x].$$
\end{exa}

Let $0 \leq i \leq n-1$ and $C_q(i,n)$ be the set defined by: $C_q(i,n)=\{i, iq, iq^2,...,i q^{m_i-1}\}$where $m_i$ is the smallest positive integer such that $iq^{m_i} \equiv i \mod n.$ This set is called the  $q$-cyclotomic coset of $n$ containing $i$. Let $I$ be a complete set of representatives of the $q$-cyclotomic cosets modulo $n$. We recall that the decomposition of $X^n- 1$ into a product of basic irreducible pairwise coprime polynomials in $R[X]$ is given by:
$X^n-1=\prod_{i \in I} f_{i}(X),$ where $f_i(X)=\prod_{j \in C_q(i,n)}(X- \xi^j),$ and $\xi$ is a primitive  $nth$-root of unity.
It is well-known that $f_i$ and $f_i^*$ are associated if and only if $ C_q(i,n) = C_q(n-i,n)$ if and only if $q^{l}\equiv -1 \mod n$ for some integer $l$ (see \cite{ref46, ref35}).

\begin{theo}\label{carcy}
Let  $\C$ be a cyclic code of length $n$ over $R$ with even index of nilpotency $t$. There exists a non-trivial self-dual code of length $n$ over $R$ if and only if $q^{i} \not\equiv -1 \mod n,$ for all positive integers $i.$
\end{theo}
\pr
Assume that there exists a non-trivial self-dual code $\C$ over $R$ such that $\C= \oplus_{i=0}^{l-1} \gamma^{r_i} \theta_i R[x],$ with
$0 \leq r_0 < r_1 <...<r_{l-1}<  t,$ then by previous theorem, there exists $\theta_i \in \{\theta_0,..., \theta_{l-1}\}$ such that $\theta_{i}$ and $\theta_i^{*}$ are not associated. We can write $\theta_i$ in the form $\theta_i= \sum \limits_{\substack{j \in J \\ J \subset I}} e_j,$ where $(e_j)_{j \in J}$ is a subset of  the complete set of primitive pairwise orthogonal idempotents of $R[X]/<X^n-1>= R[x]$. Since  $\theta_{i}$ and $\theta_i^{*}$ are not associated, then $e_j$ and $e_j^{*}$ are not associated  $\forall ~j \in J$. From Theorem \ref{joe1}, there exists $u_i \in R[x]$ such that $ e_j= u_j(x) \hat{f}_j(x)$. Then $e_j$ and $e^*_j$ are associated if and only if $\hat{f}_j$ and $\hat{f}_j^{*}$ are associated if and only if $f_j$ and $f_j^*$ are associated. But  $f_j$ and $f_j^*$  are associated if and only if $C_q(j,n)= C_q(n-j,n)$ if and only if  $q^k \equiv -1 \mod n$ for some integer $k$.

\begin{flushright}
\qed
\end{flushright}

The following result characterizes non-trivial cyclic self-dual codes over $R$ of odd or oddly even length.

\begin{theo}[\cite{ref72} Theorem 4.6]{~\newline}\label{kenz}
Let $n$ be an odd integer and $R$ be a finite chain ring with even index of nilpotency $t$. There exists non-trivial cyclic self-dual codes of length $n$ or $2n$ over $R$ if and only if the multiplicative order of $q$ modulo $n$ is odd.
\end{theo}

The following two results are consequences  of Theorem \ref{carcy} and Theorem \ref{kenz}.

\begin{pro}[\cite{ref46}, Corollary 4.6]{~\newline}
 Let $R$ be a finite chain ring with even index of nilpotency $t$ and residue field $\F_{p^r}.$ If $n$ is prime, then non-trivial self-dual codes of length $n$ do not exist in the following cases:
\begin{itemize}
\item[$\bullet$] $p=2, ~n\equiv 3,5 \mod 8;$
\item[$\bullet$] $p=3, ~n \equiv 5,7 \mod 12;$
\item[$\bullet$] $p=5, ~n \equiv 3,7,13,17 \mod 20;$
\item[$\bullet$] $p=7, ~n \equiv 5,11,13,15,17,23 \mod 28;$
\item[$\bullet$] $p=11, ~n \equiv 3,13,15,17,21,23,27,29,31,41 \mod 44.$
\end{itemize}
\end{pro}

\begin{pro}[\cite{ref72}, Corollary 4.8 and 4.9]{~\newline}
 Let  $R$ be a finite chain ring with even index of nilpotency $t$ and residue field $\F_{p^r}.$
\begin{enumerate}
 \item Let $n=\prod_{i=1}^s p_i^{k_i}$ be the prime factorization of an odd integer $n$. If $q$ is a quadratic residue of $p_i^{k_i}$ and $p_i \equiv -1 \mod 4,~ \forall~ 1 \leq i \leq s;$ then there exists a non-trivial self-dual code of length $n$ over $R$.

\item Let $n$  be an odd prime integer such that $n \equiv -1 \mod 4.$ Then there exists a non-trivial self-dual code of length $n$ over $R$ if and only if $p$ is a quadratic residue of $n^k;$ for $k$ a non-zero positive integer.
\end{enumerate}
\end{pro}

\section{Self-dual negacyclic codes}
Note that if $n$ is odd, then there exists a one-to-one correspondence between cyclic and negacyclic codes of length $n$ over $R$ (see Theorem $4.3$ in \cite{ref78} or Proposition $5.1$ in \cite{ref46}). For this reason, we only consider negacyclic codes of even length.

The following result and its proof are similar to Theorem \ref{similar}.
\begin{theo}
Under the same assumptions as the Theorem \ref{joe1}, let $\C$ be a negacyclic code of even length $n$ over $R$ with index of nilpotency $t$, such that
$$\C= \oplus_{i=0}^{l-1} \gamma^{r_i} \theta_i R[x],$$ with
$0 \leq r_0 < r_1 <...<r_l= t.$
\begin{itemize}
 \item[$i)$] If $t$ is even, there exists a non-trivial self-dual code over $R$ if and only if there exists an idempotent $\theta_i \in \{\theta_0,..., \theta_{l-1}\}$ such that $\theta_i$ and $\theta_i^*$are not associated.

\item[$ii)$] If $t$ is odd, there exists a negacyclic self-dual code over $R$ if and only if  $\theta_i$ and $\theta_i^*$ are not associated for all $\theta_i \in \{\theta_0,..., \theta_{l-1}\}$.
\end{itemize}
\end{theo}

Since $X^n+1 =(X^{2n}-1)/(X^n-1),$ then $X^n+1$ can be factored uniquely into monic irreducible pairwise coprime polynomials as follows (see \cite{ref70}):

$X^n+1=\prod_{i \in I_{2n} \cap O_{2n}} f_i(X)$ with $f_i= \prod_{i \in C_q(i,2n) \cap O_{2n}}(X- \xi^{i}_{2n}),$ where $I_{2n}$ is a complete set of representatives of cyclotomic cosets modulo $2n$, $O_{2n}$ is the set of odd integers from $1$ to $2n-1$ and $\xi_{2n}$ is a $2nth$-root of unity.\\
Similarly to Theorem  \ref{carcy}, we have the following result.

\begin{theo} \label{carne}
Let $\C$ be a cyclic code of even length $n$ over $R$. There exists a non-trivial cyclic self-dual code over $R$ if and only if $q^{i} \not\equiv -1 \mod 2n,$ for all positive integers $i.$
\end{theo}

\end{document}